\documentclass[pra,aps,10pt,twocolumn,showpacs,amsmath,amssymb]{revtex4-1}

\usepackage[english]{babel}
\usepackage{afterpage}
\usepackage{graphicx}
\usepackage{dcolumn}
\usepackage{bm}
\usepackage{color}
\usepackage{slashed}
\usepackage{latexsym}

\DeclareMathAlphabet\mathbfcal{OMS}{cmsy}{b}{n}

\newcommand{\dd}{\mathrm{d}}  
\newcommand{\ii}{\mathrm{i}}  
\newcommand{\ee}{\mathrm{e}}  

\begin{document}

\title{Threshold effects in electron-positron pair creation from the vacuum:\\ Stabilization and longitudinal vs transverse momentum sharing}
\author{K. Krajewska}
\email[E-mail address:\;]{Katarzyna.Krajewska@fuw.edu.pl}
\author{J. Z. Kami\'nski}

\affiliation{Institute of Theoretical Physics, Faculty of Physics, University of Warsaw, Pasteura 5,
02-093 Warszawa, Poland}
\date{\today}

\begin{abstract}
Momentum distributions of electron-positron pairs created from the vacuum by an oscillating in time electric field are calculated in the framework of
quantum field theory. A pronounced enhancement of those distributions is observed as the frequency of the electric field passes across the one-photon
threshold. Below that threshold the pairs preferentially carry a longitudinal momentum, while above the threshold they tend to carry a transverse momentum. 
Such momentum sharing has an impact on the number of produced pairs: It grows fast with increasing the field frequency
below the threshold but it saturates at a roughly constant value above it. On the other hand, at the fixed frequency above the one-photon threshold, 
the number of pairs scales quadratically with the field strength. This typically perturbative scaling holds even for large electric fields. Thus, 
the validity of the perturbation theory is extended here to processes which result in creation of particles with substantial transverse momenta.
\end{abstract}

\maketitle

\section{Introduction}
\label{intro}

The vacuum instability in the presence of a static electric field, which results in electron-positron ($e^-e^+$) pair creation, has been predicted decades 
ago~\cite{Sauter,Heisenberg-Euler,Schwinger}. Breaking the vacuum requires an enormous electric field strength, ${\cal E}_S=m_{\rm e}^2c^3/|e|=1.32\times 10^{18}$~V/m,
where $m_{\rm e}$ is the electron rest mass and $e=-|e|<0$ is its charge (here and in what follows, we keep $\hbar=1$)~\cite{Fradkin}. Here, ${\cal E}_S$ is the Sauter-Schwinger
critical field. Since such electric field cannot be achieved in laboratory settings, the Sauter-Schwinger mechanism of electron-positron pair production has 
not been verified experimentally yet. Another disadvantage is that the process is very weak. Hence, various
proposals have been put forward aiming at enhancing the signal of electron-positron pairs~\cite{Schutzhold,Dunne1,Orthaber,Fey,Jansen,Akal,Otto1,Otto2,Jiang,Sitiwaldi,
Akkermans,Li1,Li2,Dumlu1,Dumlu2,Xie,KTK}. Thus, raising the question of optimal control of the process~\cite{Hebenstreit,Fillion}.

In this paper, we study the Sauter-Schwinger pair production by a pulsed electric field, with a frequency tuned so it passes 
across the one-photon threshold. This region is particularly interesting, due to a threshold-related enhancement of probability distributions of created pairs.
Similar enhancements have been observed before in the context of multiphoton pair production and channel closing effects~\cite{CC2,CC1,CC3}.
It is, however, the one-photon process that is the most efficient of all, in agreement with the results presented in~\cite{CC1,CC3}. 
At this point, let us also note that similar threshold-related enhancements have been observed in other multiphoton processes such as 
strong-field detachment or ionization~\cite{XX,AA,YY,ZZ} and, as such, they are a universal strong-field phenomenon.

A closely related to threshold enhancements is the concept of an effective mass acquired by particles in strong fields. Such a concept, depending explicitly 
on the electric field parameters, has been discussed in~\cite{CC1}. Its limitation was also noted, as the effective mass should vary between different parameter
regimes. While~\cite{CC1,CC3} are related to the nonperturbative regime of pair creation, most of our results concern the opposite one. It is commonly believed that 
in the perturbative regime, the effective mass approaches the rest mass of particles. As we demonstrate in our paper, this is provided that the transverse momentum 
of created particles is negligible.

In the current paper, we provide a detailed study of the Sauter-Schwinger process in the vicinity of the one-photon threshold.
This threshold marks the border between a multiple- and one-photon pair production. Thus, it makes for a qualitatively different behavior
of the resulting momentum distributions and the number of produced pairs in those two regimes. Our analysis shows, for instance, that
the process is either dominated by the longitudinal or by the transverse motion of created particles, that is  below and above the one-photon threshold respectively.
This also affects the marginal momentum distributions of created particles, as presented in the paper. The overall result on the total number of produced pairs is
that while it increases fast with the electric field frequency below the threshold, it stabilizes above it. While our conclusions follow
from exact numerical calculations, they are also confirmed by analytical results derived from the perturbative treatment of pair production.

Note that stabilization by external fields has been observed in other quantum-mechanical problems as well.
See, for instance, the stabilization of atoms or molecules in intense laser fields~\cite{Gav} and the stabilization of electron states in 
semiconductor heterostructures by crossed magnetic and electric fields~\cite{KK}. Such an effect has also been observed in the nonlinear Bethe-Heitler process
of pair production~\cite{KKK}. This counter-intuitive effect seems, therefore, to occur quite frequently in quantum physics.

The paper is organized as follows. In Sec.~\ref{theory}, for convenience of the reader, we briefly present the theoretical formulation of the Sauter-Schwinger 
pair production by a time-dependent electric field. This is followed by a detailed analysis of the resulting momentum distribution and the number of created pairs 
in Sec.~\ref{numerics}. We summarize our results in Sec.~\ref{summary}.


\section{Theoretical formulation}
\label{theory}

We consider the electron-positron pair creation from vacuum by a homogeneous in space, time-dependent electric field, ${\mathbfcal{E}}(t)={\cal E}(t){\bm e}_z$.
In addition, we assume that the field is generated by lasers, meaning that the condition,
\begin{equation}
\int_{-\infty}^{+\infty}\dd t\,{\cal E}(t)=0,
\label{el1}
\end{equation}
is satisfied~\cite{KTK}. In the presence of the time-dependent electric field, the fermionic field operator can be decomposed into eigenmodes which are labeled 
by the linear momentum ${\bm p}$. In the following, we will relate to its longitudinal $p_\|$ and transverse ${\bm p}_\perp$ components being defined with respect 
to the electric field direction, 
\begin{equation}
p_\|={\bm p}\cdot {\bm e}_z,\quad {\bm p}_\perp={\bm p}-p_\| {\bm e}_z.\label{el2}
\end{equation}
As it was demonstrated in~\cite{Grib1,Grib2,KTK}, the momentum distribution of particles $f({\bm p})$ generated in the given eigenmode of the fermionic field ${\bm p}$ is
determined by solving the system of equations,
\begin{equation}
\ii\frac{\dd}{\dd t}\begin{bmatrix}
                     c_{\bm p}^{(1)}(t)\\
                     c_{\bm p}^{(2)}(t)\\
                     A(t)
                    \end{bmatrix}
                    =
                    \begin{pmatrix} \omega_{\bm p}(t) & \ii\Omega_{\bm p}(t) & 0 \cr -\ii\Omega_{\bm p}(t) & -\omega_{\bm p}(t) & 0 \cr 0 & 0 & 0\end{pmatrix}
                    \begin{bmatrix}
                     c_{\bm p}^{(1)}(t)\\
                     c_{\bm p}^{(2)}(t)\\
                     A(t)
                    \end{bmatrix}
                    -\begin{bmatrix}
                     0\\
                     0\\
                     \ii {\cal E}(t)
                    \end{bmatrix},
\label{a3}
\end{equation}
with the initial conditions that 
\begin{equation}
\lim_{t\rightarrow -\infty}c_{\bm p}^{(1)}(t)=1, \quad \lim_{t\rightarrow -\infty} c_{\bm p}^{(2)}(t)=0, \quad \lim_{t\rightarrow -\infty} {\cal A}(t)=0.
\label{initial}
\end{equation}
Namely,
\begin{equation}
f({\bm p})=\lim_{t\rightarrow +\infty}|c_{\bm p}^{(2)}(t)|^2,\label{el8}
\end{equation}
where
\begin{equation}
|c_{\bm p}^{(1)}(t)|^2+|c_{\bm p}^{(2)}(t)|^2=1.\label{el9}
\end{equation}
Moreover, $A(t)$ in~\eqref{a3} defines the vector potential, ${\bm A}(t)=A(t){\bm e}_z$, where ${\cal E}(t)=-\dot{A}(t)$, and, according to Eq.~\eqref{el1}, it has to satisfy the condition,
\begin{equation}
\underset{t\rightarrow-\infty}{\lim}A(t)=\underset{t\rightarrow+\infty}{\lim}A(t)=0.
\label{a4}
\end{equation}
Here, we took into account Eq.~\eqref{initial}. Moreover, in Eq.~\eqref{a3}, we have introduced,
\begin{equation}
\omega_{\bm p}(t)=\sqrt{(m_{\rm e}({\bm p}_\perp)c^2)^2+c^2(p_\|-eA(t))^2},\label{a5}
\end{equation}
with the electron effective mass,
\begin{equation}
m_{\rm e}({\bm p}_\perp)=\frac{1}{c}\sqrt{(m_{\rm e}c)^2+{\bm p}_\perp^2}.\label{a7}
\end{equation}
This is to emphasize that the transverse momentum of the electron always enters the formulas through the coupling to $m_{\rm e}c$. This is confirmed
by the definition of
\begin{equation}
\Omega_{\bm p}(t)=-ce{\cal E}(t)\frac{m_{\rm e}({\bm p}_\perp)c^2}{2\omega_{\bm p}^2(t)},\label{a6}
\end{equation}
which appears in~\eqref{a3} along with $\omega_{\bm p}(t)$. Note that, for negligibly small transverse momenta of produced particles, their effective mass~\eqref{a7}
reduces to the rest one. As we will show in the next section, this characterizes the perturbative regime of a few-photon pair production.
In contrast, the one-photon processes are typically accompanied by a large transverse momentum gain. Thus, giving a rise to the particles rest mass.

\begin{figure}
\centering
\includegraphics[width=0.4\textwidth]{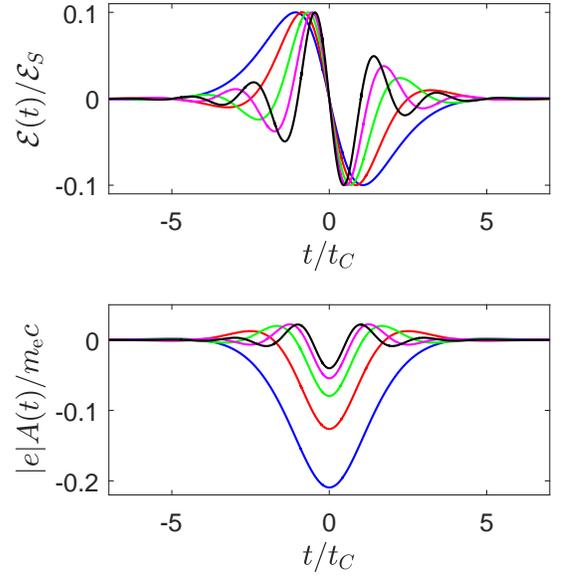}
\caption{Time dependence of the electric field $\mathcal{E}(t)$ [Eqs.~\eqref{el15} and~\eqref{el16}] and the corresponding vector potential $A(t)$ for different values of the parameter $N=$ 1 
(blue), 2 (red), 3 (green), 4 (magenta), and 5 (black). With increasing $N$ the carrier frequency of the electric field oscillations $\omega$ increases as well. 
On the other hand, we keep the amplitude of the electric field fixed, ${\cal E}_0=-0.1{\cal E}_S$. Hence, while the number of modulations in the pulse increases with $N$,
the amplitude of the vector potential decreases like $1/N$. 
}
\label{fig1}
\end{figure}

The first two equations of~\eqref{a3} are structurally identical with the Schr\"odinger equation of a two-level system. They have to be solved for the functions 
$c_{\bm p}^{(i)}(t)$, $i=1,2$, which determine the momentum distributions of created particles [Eqs.~\eqref{el8} and~\eqref{el9}]. 
Due to the axial symmetry of the problem, those distributions will depend only on $p_\|$ and $p_\perp^2={\bm p}^2-p_\|^2$,
i.e., $f({\bm p})=f(p_\|,p_\perp)$. For our further purposes, we also introduce the marginal momentum distributions,
\begin{equation}
f(p_{\|})=2\pi\int_0^{+\infty} \dd p_{\bot}\, p_{\bot} f(p_{\|},p_{\bot})\label{el10}
\end{equation}
and
\begin{equation}
f(p_{\bot}^2)=\pi\int_{-\infty}^{+\infty}\dd p_{\|}\, f(p_{\|},p_{\bot}).\label{el11}
\end{equation}
Then, the total number of pairs created in the relativistic unit volume becomes
\begin{align}
f&=\int \dd^3 p f(p_{\|},p_{\bot})=2\pi\int_{-\infty}^{+\infty}\dd p_{\|} \int_0^{+\infty} p_{\bot}\dd p_{\bot}\, f(p_{\|},p_{\bot}) \nonumber\\
&=\int_{-\infty}^{+\infty} \dd\, p_{\|} f(p_{\|}) =\int_0^{+\infty} \dd p_{\bot}^2\, f(p_{\bot}^2). \label{el12}
\end{align}
Note that these definitions allow us to determine the momentum probability distributions,
\begin{equation}
\mathcal{P}(p_{\|},p_{\bot})=f(p_{\|},p_{\bot})/f\label{el13}
\end{equation}
and
\begin{equation}
\mathcal{P}(p_{\|})=f(p_{\|})/f,\quad
\mathcal{P}(p_{\bot}^2)=f(p_{\bot}^2)/f.\label{el14}
\end{equation}
While in the following we will calculate $f(p_\|,p_\perp)$ along with the longitudinal and transverse momentum distributions of created particles, $f(p_\|)$ and 
$f(p_\perp^2)$, their functional dependence on $p_\|$ and $p_\perp^2$ remains the same as for the probability distributions mentioned above. Thus, it will be justified
to talk about the particle's momenta for which the pair creation will be the most/least probable, even though formally we will not calculate the probability distributions
given by Eqs.~\eqref{el13} and~\eqref{el14}.

We will perform calculations for the following model of a pulsed electric field, 
\begin{equation}
\mathcal{E}(t)=\mathcal{E}_0 F(t),\label{el15}
\end{equation}
with
\begin{equation}
F(t)=\frac{{\cal N}_0}{\cosh(\beta t)}\sin(\omega t).\label{el16}
\end{equation}
Here, ${\cal E}_0$ is the amplitude while $\omega$ is the carrier frequency of field oscillations. The parameter $\beta$ in~\eqref{el16} determines the bandwidth of the pulse.
In the following, we will fix $\beta=m_{\rm e}c^2$ and ${\cal E}_0$ to either ${\cal E}_0=-0.1{\cal E}_S$ or ${\cal E}_0=-{\cal E}_S$. On the other hand, we will change the frequency 
$\omega$ such that $\omega=N\omega_0$, with $\omega_0=0.2\pi\,m_{\rm e}c^2$ and the value of $N$ varying continuously. To fix the amplitude of the electric field while changing its carrier frequency, we will adjust 
${\cal N}_0$ such that
\begin{equation}
\underset{t}{\max} |F(t)| =1.\label{el17}
\end{equation}
This means that the number ${\cal N}_0$ has to be changed appropriately for different values of $N$. For ${\cal E}_0=-0.1{\cal E}_S$, the time dependence of the electric field 
${\cal E}(t)$ and the corresponding vector potential $A(t)$ are shown in Fig.~\ref{fig1}. Note that the maximum of the vector potential scales like $1/N$, 
and so it decreases with increasing $N$. Similar behavior of ${\cal E}(t)$ and $A(t)$ is observed if we change the electric
field amplitude to ${\cal E}_0=-{\cal E}_S$. As we will demonstrate in the next section, this will have an impact on the momentum distributions of created pairs.

\section{Numerical illustrations}
\label{numerics}
\begin{figure}
\centering
\includegraphics[width=0.4\textwidth]{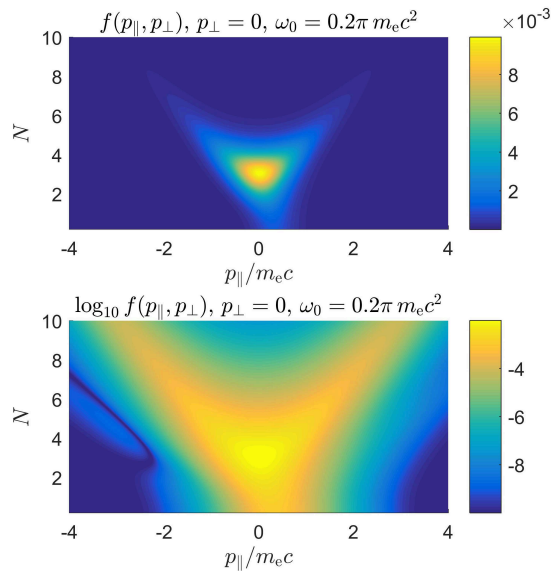}
\caption{Color mappings of the momentum distribution of created particles, $f(p_{\|},p_{\bot})$, for $\mathcal{E}_0=-0.1\mathcal{E}_S$ and ${\bm p}_{\bot}={\bm 0}$ as a function of $p_{\|}$ 
and $N=\omega/\omega_0$. While in the upper panel we plot the results in the linear scale, in order to emphasize the details of the distribution the same but in the logarithmic 
scale is plotted in the lower panel. The distribution is maximum roughly for $p_{\|}\approx 0$ and $\omega\approx 2m_{\mathrm{e}}c^2$ which, together with ${\bm p}_{\bot}={\bm 0}$,
define the threshold for the one-photon pair production.
}
\label{fig2}
\end{figure}
\begin{figure}
\centering
\includegraphics[width=0.4\textwidth]{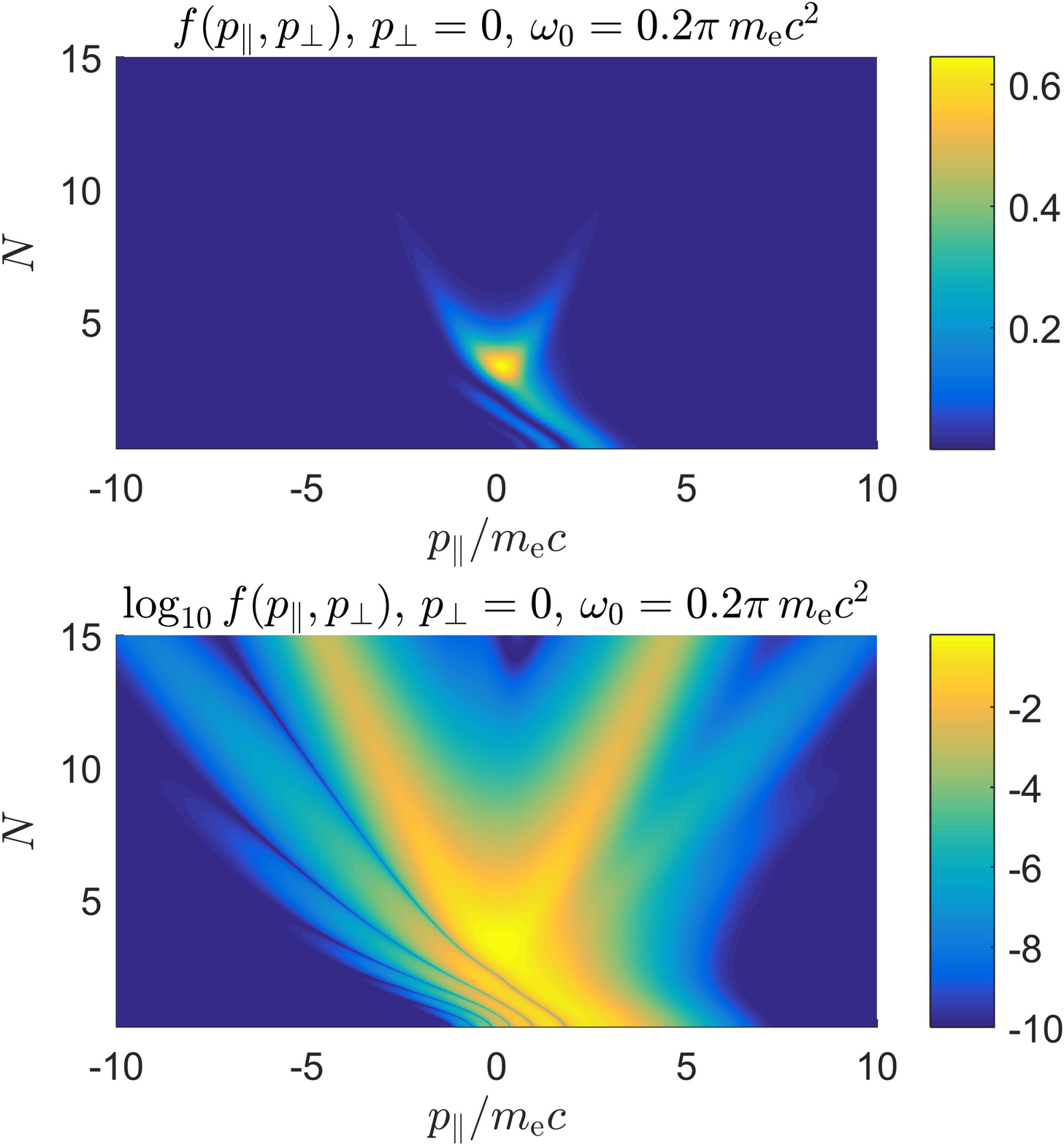}
\caption{The same as in Fig.~\ref{fig2} but for  $\mathcal{E}_0=-\mathcal{E}_S$. 
}
\label{fig3}
\end{figure}

\begin{figure*}
\centering
\includegraphics[width=0.7\textwidth]{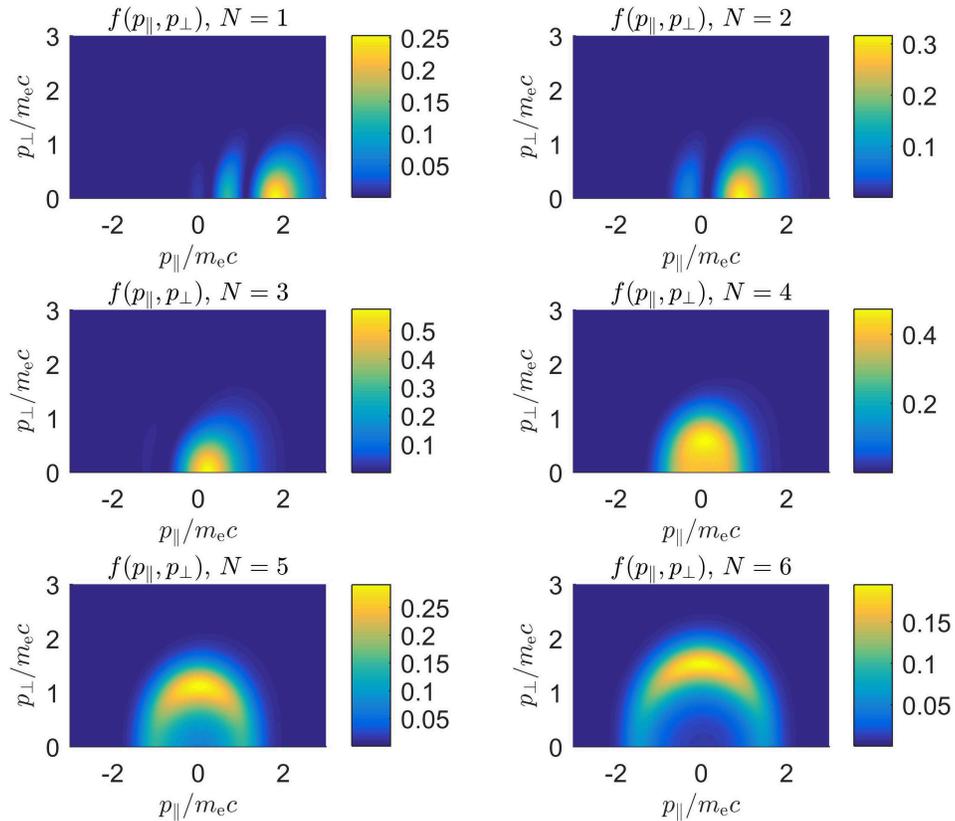}
\caption{Color mappings of the momentum distribution of created particles, $f(p_{\|},p_{\bot})$, as a function of their longitudinal and transverse momenta. 
The results (in the linear scale) are plotted for $\mathcal{E}_0=-\mathcal{E}_S$ and for different values of $N$, as denoted in each panel. For $N=1,2$, and $3$, 
the process occurs via absorption of multiple electric field quanta. In this case, the most probable is to produce the $e^-e^+$ pair with a roughly
zero transverse and nonzero longitudinal momentum components. With increasing $N$ ($N\geqslant 4$), the pair creation occurs via a one-photon
process. In such case, the maximum of the distribution shifts toward larger values of the transverse momentum. At the same time,
the distribution remains maximum around the zero longitudinal momentum, even though it becomes increasingly more spread around that value. 
Above the threshold most of pairs are created perpendicularly to the electric field.}
\label{fig4}
\end{figure*}
\begin{figure*}
\centering
\includegraphics[width=0.7\textwidth]{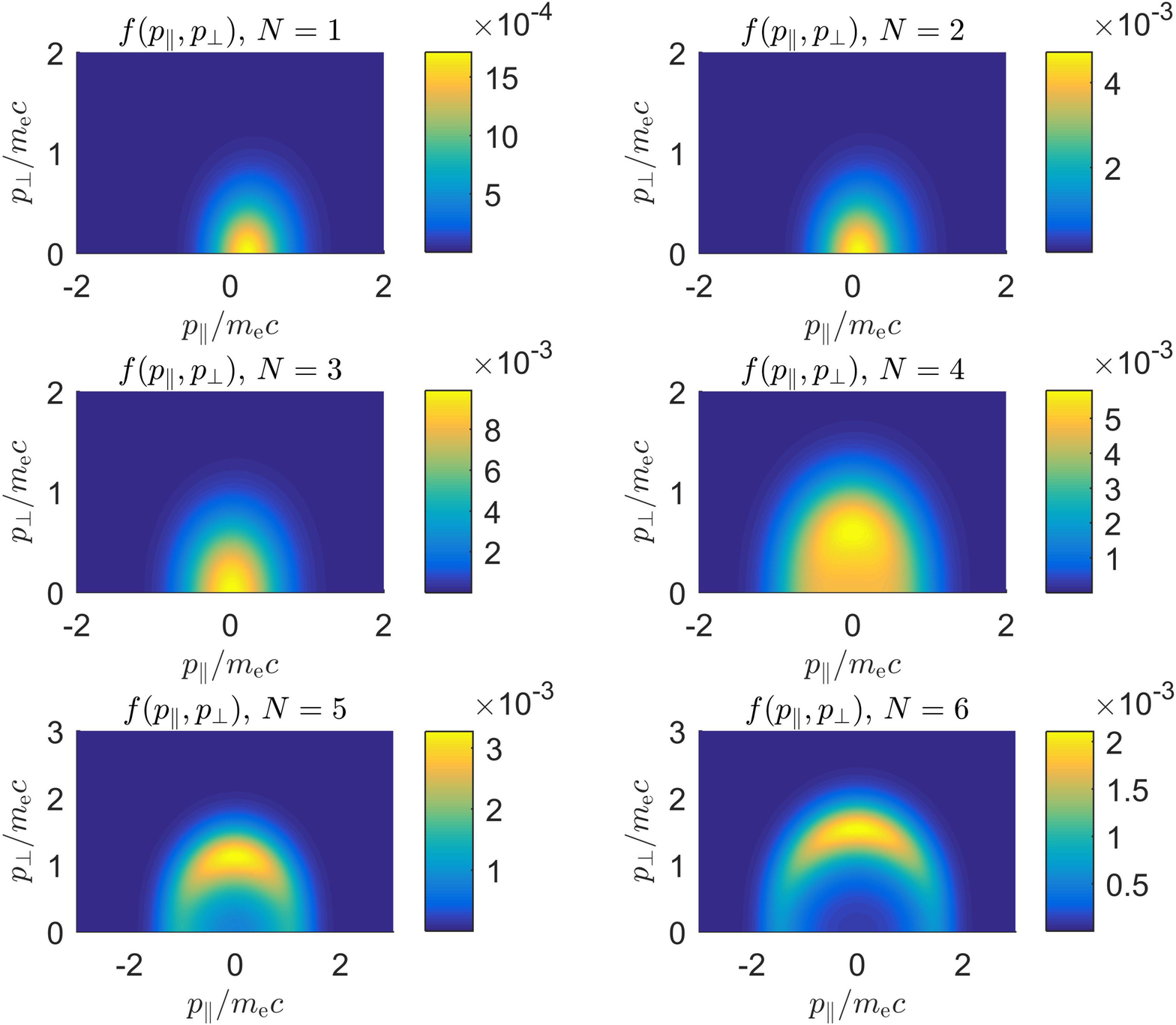}
\caption{The same as in Fig.~\ref{fig4} but for ${\cal E}_0=-0.1{\cal E}_S$.}
\label{fig4diff}
\end{figure*}

In Fig.~\ref{fig2}, we present the color mappings of the momentum distribution $f(p_\|,p_\perp)$ as a function of the longitudinal momentum $p_\|$ and a real number
$N=\omega/\omega_0$. The results are for ${\cal E}_0=-0.1{\cal E}_S$ and ${\bm p}_\perp={\bm 0}$, with the remaining parameters given in the previous section. While in the
upper panel we plot the results in the linear scale, their details become more visible in the logarithmic scale, which is used in the lower panel. As one can see, there is
a region in the $(p_\|,N)$-plane where the $e^-e^+$ pair creation is most probable. This region is characterized by a roughly zero longitudinal momentum of created pairs, 
$p_\|\approx 0$, and by the value $N\approx 3$. For $\omega_0=0.2\pi\,m_{\rm e}c^2$, the latter corresponds to the electric field carrier frequency of
roughly $\omega\approx 2m_{\rm e}c^2$. This is the energy necessary to be absorbed from the external field in order to produce an electron-positron pair 
at rest. This agrees with the fact that the corresponding momentum distribution peaks at ${\bm p}={\bm 0}$. It also shows that the most probable 
process is via absorption of a single photon. Other order processes, i.e., processes which occur via absorption of multiple photons, are possible but they are less likely to happen. 
This is even more clear in Fig.~\ref{fig3} for ${\cal E}_0=-{\cal E}_S$. The additional stripes in the distribution, which are visible already in the linear scale in the upper panel, 
correspond to different order processes. In these cases, the carrier frequency of the electric field is smaller than $2m_{\rm e}c^2$ and, hence, the one-photon pair
production is forbidden. Nevertheless, it is possible to absorb additional photons from the field, which leads to the above-threshold pair production. With increasing
$\omega$ (or, equivalently, $N$) we pass through the one-photon threshold at $\omega_{\rm th}= 2m_{\rm e}c^2$. Since for the $n$-photon process the energy conservation law
is $n\omega=2\sqrt{(m_{\rm e}({\bm p}_\perp)c^2)^2+c^2p_\|^2}$, one can anticipate that in the current case ($n=1$ and $\omega\approx \omega_{\rm th}$) it is energetically preferable 
to produce particles at rest. With still increasing $\omega$, there will be a portion of energy in excess to the threshold energy $\omega_{\rm th}$ that will be available
to the particles. As it is already indicated by Figs.~\ref{fig2} and~\ref{fig3}, which are for ${\bm p}_\perp={\bm 0}$, this portion of energy, which will be equally 
redistributed between an electron and a positron, is hardly converted into
their longitudinal motion. A more detailed analysis of the excess energy sharing between the longitudinal and the transverse motion of the particles can be based
upon the color mappings presented in Fig.~\ref{fig4}.


The two-dimensional momentum mappings in Fig.~\ref{fig4} are for ${\cal E}_0=-{\cal E}_S$ and for different values of $N$ (i.e., for different 
carrier frequencies $\omega$). Such strong electric field has been chosen, as the features of the momentum distributions we want to discuss next are very pronounced
in this case. First, we note that for $N=1, 2,$ and $3$, the corresponding frequencies of the field oscillations, $\omega=N\omega_0$, 
are below the one-photon energy threshold $\omega_{\rm th}$. In other words, these processes correspond to the above-threshold pair production. In this regime,
the pairs are mostly created with a zero transverse and a nonzero longitudinal momentum components. For this strong field, already in the linear scale, the multiphoton 
modulations in the pair creation momentum distributions are visible. More importantly, with increasing $N$ toward the one-photon threshold ($N_{\rm th}=10/\pi=3.18$)
the maximum of those distributions shifts toward zero longitudinal momentum. At the same time, when looking at the magnitude of the distributions,
those will be the most pronounced of all. While still increasing $N$, the pair creation remains the one-photon process. Such a process typically occurs with
a nearly zero longitudinal momentum of created particles but the excess energy absorbed from the field is converted into the particles transverse motion. 
Thus, with increasing $N$ (and, hence, with increasing the available excess energy), the maximum of the momentum distribution is shifted toward larger transverse
momenta.

As it follows from our analysis, the energy sharing between the transverse and longitudinal motion of created particles strongly depends on whether the pair creation
is due to a single- or to a multi-photon transition. In either case, the energy absorbed from the field seems to be redistributed such 
that the particles have minimal energy. Specifically, for small $N$, in which case the vector
potential takes large values (see, Fig.~\ref{fig1}), it has to be balanced by a large longitudinal momentum. This follows from the definition of $\omega_{\bm p}(t)$
[Eq.~\eqref{a5}] which enters the system of equations~\eqref{a3}. It also explains why with
decreasing $N$, the respective longitudinal momentum for which we observe maximum pair creation shifts toward larger values. At the same time, to minimalize the energy
of created particles, the transverse momentum remains roughly zero. Taking into account the definition~\eqref{a7}, one can conclude that multi-photon pair creation
occurs with no increase of the particles effective mass. According to Eq.~\eqref{a5}, the resulting momentum distributions should not be symmetric
with respect to the momentum reflection $p_\|\rightarrow -p_\|$, or, consequently, ${\bm p}\rightarrow -{\bm p}$. This is confirmed by Fig.~\ref{fig4} for substantial
values of the vector potential, i.e., for small values of $N$. However, for larger $N$ and, hence, for smaller values of the vector potential describing the external
field, this asymmetry vanishes. As already discussed, close to the one-photon threshold, the pairs are created with roughly a zero momentum, ${\bm p}\approx {\bm 0}$.
With still increasing $N$, the coupling to the vector potential practically does not play a role. In this case, it is energetically preferable that the excess energy 
absorbed from the field contributes to the transverse motion of created particles. Hence, the one-photon pair creation typically is accompanied by an increase
of the particles effective mass~\eqref{a7}.


To illustrate the transition between the pair creation while it occurs with multiple vs. single photon absorption we have used a rather strong electric field,
${\cal E}_0=-{\cal E}_S$. For weaker fields, such as ${\cal E}_0=-0.1{\cal E}_S$, the general features of the momentum distributions remain however
the same (see, Fig.~\ref{fig4diff}). The difference is that in this case the multiphoton modulations of the momentum distributions are not as pronounced as in Fig.~\ref{fig4}.
This will affect detailed features of the marginal momentum distributions presented below.

In Fig.~\ref{fig5}, we show the transverse momentum distribution $f(p_\perp^2)$ of pairs generated from the vacuum 
by the electric field considered in this paper. This time, the amplitude of the electric field oscillations is ${\cal E}_0=-0.1{\cal E}_S$. Each curve 
corresponds to a different carrier frequency of the electric pulse determined by $N$. As the effect of the integration of the two-dimensional 
momentum maps with respect to the longitudinal momentum component, for those frequencies which are below the one-photon energy threshold, $f(p_\perp^2)$ 
has a maximum at the zero perpendicular momentum. With increasing $N$, the maximum of the distribution grows in magnitude
to become the most pronounced at the threshold value $N_{\rm th}=10/\pi$. Above the threshold, however, which is illustrated in Fig.~\ref{fig5} for $N=4,5,$ and $6$, the 
distribution exhibits an off-axis maximum. With increasing $N$, this maximum shifts toward larger values of $p_\perp$. While it spreads in $p_\perp$,
the magnitude of the distribution drops down. The exact same behavior of $f(p_\perp^2)$ is observed for stronger electric fields. Interestingly, 
as we have checked this for ${\cal E}_0=-{\cal E}_S$, the off-axis maxima of the transverse momentum distributions occur at the exact same values 
of $p_\perp$ for the given $N$. The respective values of $p_\perp$ can be estimated from the energy conservation relation. Namely, taking into account that the pairs 
produced in the one-photon process have most likely a zero longitudinal momentum, we estimate that $p_\perp=\sqrt{(N\omega_0/2c)^2-(m_{\rm e}c)^2}$.
This predicts quite accurately the positions of the off-axis maxima in Fig.~\ref{fig5} and is independent of the amplitude of the electric field oscillations ${\cal E}_0$.

In Fig.~\ref{fig6}, we plot the longitudinal momentum distribution $f(p_\|)$ (upper panel) and the total number of created pairs $f$ per unit volume (lower panel) for
${\cal E}_0=-0.1{\cal E}_S$. While all curves in the upper panel are bell-shaped, their maximum is shifted to positive values of $p_\|$ for small $N$. 
This asymmetry is lifted with increasing $N$, which was already noted in relation to the two-dimensional distributions $f(p_\|,p_\perp)$. If we perform similar 
calculations for ${\cal E}_0=-{\cal E}_S$, the bell-shaped curves centered at $p_\|\approx 0$ represent the results for $N>N_{\rm th}$ only. For small $N$,
the marginal momentum distribution $f(p_\|)$ exhibits multiple maxima for $p_\|>0$, which result from the multiphoton modulations observed in Fig.~\ref{fig4}. Rather than that, 
the overall behavior of $f(p_\|)$ is the same, irrespectively of the electric field strength. In the lower panel of Fig.~\ref{fig6}, we present the total number of pairs $f$ 
calculated for discrete values of $N$ between 1 and 20. Below the one-photon-threshold ($N<N_{\rm th}$), $f$ takes rather small values which, however, increase quickly around $N_{\rm th}$.
Above the one-photon threshold ($N>N_{\rm th}$), on the other hand, the number of created particles seems to saturate and even slightly decreases with $N$.
One can conclude that, while the excess energy absorbed from the field in the one-photon process
increases the effective mass of created particles $m_{\rm e}({\bm p}_\perp)$, it hardly affects their number.
Similar stabilization is observed for ${\cal E}_0=-{\cal E}_S$, even though the number of pairs in this case is by two orders of magnitude larger. 
In light of this result, it is interesting to study more closely the dependence of
the number of created particles on the electric field strength.

\begin{figure}
\centering
\includegraphics[width=0.4\textwidth]{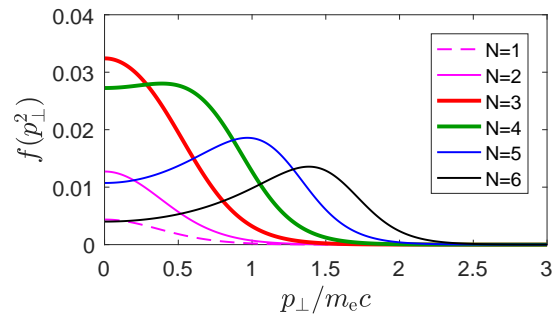}
\caption{The transverse momentum distribution $f(p_\perp^2)$ for ${\cal E}_0=-0.1{\cal E}_S$ while changing the carrier frequency of the electric field, as
pointed out by different values of $N$. The thick lines correspond to processes either just below ($N=3$) or above one-photon threshold ($N=4$).}
\label{fig5}
\end{figure}

\begin{figure}
\centering
\includegraphics[width=0.4\textwidth]{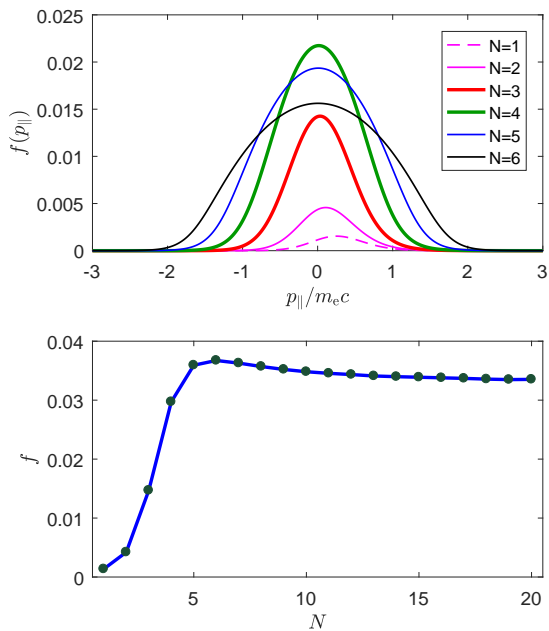}
\caption{The longitudinal momentum distribution of created pairs $f(p_{\|})$ (upper panel) and the total number of pairs created in the relativistic unit volume $f$ (lower panel)
for $\mathcal{E}_0=-0.1\mathcal{E}_S$ and for chosen values of, in principle, real parameter $N$. Similar to Fig.~\ref{fig5}, the thick lines correspond to just below ($N=3$) 
or just above one-photon threshold ($N=4$).
}
\label{fig6}
\end{figure}

\begin{figure}
\centering
\includegraphics[width=0.4\textwidth]{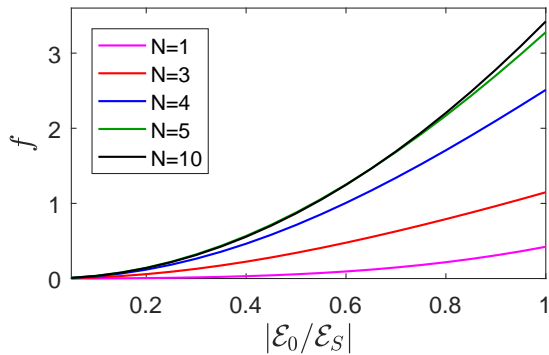}
\caption{Dependence of the total number of $e^-e^+$ pairs, $f$, on the electric field strength ${\cal E}_0$. 
Each curve corresponds to a different value of $N$, which translates into a different carrier frequency of the external field according to $\omega=N\omega_0$.
}
\label{fig7}
\end{figure}

In Fig.~\ref{fig7}, we plot the total number of created pairs $f$ [Eq.~\eqref{el12}] as a function of ${\cal E}_0$. The results are for different $N$, as denoted in the figure. 
We have checked that, above the one-photon-threshold, the curves fit well to the functional dependence: $f=3.5({\cal E}_0/{\cal E}_S)^2$. In other words, 
for $N>N_{\rm th}$, the number of created pairs per relativistic unit volume depends quadratically on the external field strength, as observed already in relation to Fig.~\ref{fig6}.
This indicates the perturbative character of the process. The same conclusion is reached when considering the Keldysh parameter $\gamma=\omega {\cal E}_S/(m_{\rm e}c^2|{\cal E}_0|)$. Specifically, 
we would also like to note that $\gamma=2\pi N$ for ${\cal E}_0=-0.1{\cal E}_S$ and $\gamma=\pi N/5$ for ${\cal E}_0=-{\cal E}_S$, which are the cases thoroughly studied in this paper. 
This shows that most of our results relate to the perturbative regime of electron-positron pair creation, with $\gamma\gtrsim 1$. Let us also note that, for fixed ${\cal E}_0$, 
we go even more deeply into this regime when increasing $N$. Going back to Fig.~\ref{fig7}, we note that neither 
of the curves can be fitted to the Schwinger tunneling formula $f\sim ({\cal E}_0/{\cal E}_S)^2\exp\bigl(-\pi {\cal E}_S/|{\cal E}_0|\bigr)$, as
the latter is applicable for $\gamma\ll 1$.

In light of these conclusions, let us go back to Eq.~\eqref{a3} and treat the first two equations perturbatively. This can be done assuming that
\begin{equation}
\underset{t}{\rm max}\,|\Omega_{\bm p}(t)|\ll\underset{t}{\rm min}\,|\omega_{\bm p}(t)|,\label{applicability}
\end{equation}
which means that
\begin{equation}
\frac{{\cal E}_0}{2{\cal E}_S}\ll \Bigl[\frac{m_{\rm e}({\bm p}_\perp)}{m_{\rm e}}\Bigr]^2.\label{aaa}
\end{equation}
Therefore, the perturbation theory can be applied when either the electric field is weak compared to the Sauter-Schwinger critical field, 
or when the transverse momentum of created particles is substantial (i.e., $m_{\rm e}({\bm p}_\perp)\gg m_{\rm e}$). The latter is particularly important from the point of view of this paper, as it characterizes
the one-photon processes. As we have shown numerically, beyond the one-photon threshold, the particles will be produced with a significant transverse 
momentum irrespectively of the electric field strength. This justifies the perturbative treatment of such processes, in agreement with~\eqref{aaa}.
In this case, the original system of equation~\eqref{a3} reduces to
\begin{align}
\ii\dot{c}_{\bm p}^{(1)}(t)&=\omega_{\bm p}(t)c_{\bm p}^{(1)}(t),\\
\ii\dot{c}_{\bm p}^{(2)}(t)&=-\ii\Omega_{\bm p}(t)c_{\bm p}^{(1)}(t)-\omega_{\bm p}(t)c_{\bm p}^{(2)}(t),
\end{align}
and it can be solved analytically. Accounting for the initial conditions~\eqref{initial}, we obtain
that in the lowest order perturbation theory with respect to $\Omega_{\bm p}(t)$, the momentum distribution of created pairs~\eqref{el8} becomes
\begin{equation}
f({\bm p})\approx \Bigl|\int_{-\infty}^\infty\dd t\,\Omega_{\bm p}(t)\ee^{-2\ii\int_{-\infty}^t\dd\tau\,\omega_{\bm p}(\tau)}\Bigr|^2.\label{pert}
\end{equation}
Note that the same formula can be derived using the quantum kinetic approach and the low-density approximation~\cite{Schmidt,Grib2}.
Taking into account the definition of $\Omega_{\bm p}(t)$ [Eq.~\eqref{a6}], it becomes clear that the minima of $\omega_{\bm p}(t)$ contribute
the most to the above integral. Those minima are determined by the conditions: ${\bm p}_\perp={\bm 0}$ and $p_\|= eA(t)$. The latter means that the
temporal longitudinal momentum of created particles oscillates in time following the vector potential and, hence, asymptotically it becomes zero.
A trivial conclusion is that the most probable process would result in generation of particles at rest, which can be accomplished by absorbing a photon
of energy $\omega_{\rm th}=2m_{\rm e}c^2$. This has been seen in our numerical results. In such case, $\Omega_{\bm p}(t)= -e{\cal E}_0 F(t)/(2m_{\rm e}c)$ and, consequently, the momentum
distribution~\eqref{pert} scales like ${\cal E}_0^2$. Going beyond the one-photon threshold, one can use the same argument.
The difference is that there will be the excess energy absorbed from the field that will contribute to the effective mass of particles.
In such case, $\Omega_{\bm p}(t)= -e{\cal E}_0 F(t)/(2m_{\rm e}({\bm p}_\perp)c)$, which surely indicates a quadratic scaling of the number of produced pairs
$f$ [Eq.~\eqref{pert}] with the electric field strength ${\cal E}_0$. 


\section{Summary}
\label{summary}

Our results for the momentum distributions and the number of electron-positron pairs created in the dynamical Sauter-Schwinger process 
show dramatic changes with varying the frequency of the electric field oscillations in the vicinity of the one-photon threshold. Specifically, this concerns
the energy sharing between the longitudinal and the transverse motion of created particles. While below the one-photon threshold the particles are created
with a nearly zero transverse- and a substantial longitudinal momentum, this tendency is reversed above the one-photon threshold. The latter is particularly
surprising, as it is commonly believed that particles are mostly created along the electric field. As we have shown, with increasing the field
frequency above the one-photon threshold, the field effect is exclusively to increase the effective mass of produced
particles~\eqref{a7}. Their total number, however, remains roughly the same. Moreover, it scales quadratically with the strength of the electric field,
which is typical for the perturbative regime of $e^-e^+$ pair creation. Note that such perturbative scaling of the number of produced pairs
with the electric field strength  has been demonstrated in this paper for arbitrarily strong electric fields. Thus, the 
validity of the perturbation theory has been extended in this paper to processes with substantial transverse momenta of created particles.

Our results are in line with the previous claim that the effective mass of produced particles has to be redefined in different regimes of pair creation~\cite{CC1}.
In this paper, we have proposed the effective mass description that is applicable in the perturbative regime. This involves the transverse momentum
of created particles~\eqref{a7}. While below the one-photon threshold it approaches the rest mass, it can be substantially larger than that 
above the one-photon threshold. One can conclude, therefore, that the transverse momentum of particles can be a direct signature of their effective mass. 

\section*{Acknowledgments}
This work is supported by the National Science Centre (Poland) under Grant No. 2014/15/B/ST2/02203.

\end{document}